\documentclass[aps,prl,reprint,superscriptaddress,amsmath,amssymb]{revtex4-2}

\usepackage{graphicx}
\usepackage{dcolumn}
\usepackage{bm}
\usepackage[utf8]{inputenc}
\usepackage{siunitx}
\usepackage{hyperref}

\begin{document}

\title{High-$Q$ Superconducting Quarter-Wave Resonator for Cavity Levitodynamics}

\author{Nabin Raut}
\affiliation{SRF Institute, Thomas Jefferson National Accelerator Facility, Newport News, Virginia 23606, USA.}
\author{Uttar Pudasaini}
\affiliation{SRF Institute, Thomas Jefferson National Accelerator Facility, Newport News, Virginia 23606, USA.}
\author{Eric Lechner}
\affiliation{SRF Institute, Thomas Jefferson National Accelerator Facility, Newport News, Virginia 23606, USA.}
\author{Pashupati Dhakal}
\affiliation{SRF Institute, Thomas Jefferson National Accelerator Facility, Newport News, Virginia 23606, USA.}
\author{Peter Owen}
\affiliation{SRF Institute, Thomas Jefferson National Accelerator Facility, Newport News, Virginia 23606, USA.}
\author{Robert Rimmer}
\affiliation{SRF Institute, Thomas Jefferson National Accelerator Facility, Newport News, Virginia 23606, USA.}
\author{Jay Sharping}
\affiliation{Department of Physics, University of California Merced, Merced, California 95343, USA.}
\author{Michael E. Tobar}
\affiliation{Quantum Technologies and Dark Matter Research Lab, Department of Physics, The University of Western Australia, Crawley, WA 6009, Australia.}
\author{Andrea Vinante}
\affiliation{Istituto di Fotonica e Nanotecnologie, IFN-CNR, Trento, Italy.}
\author{Anne-Marie Valente-Feliciano}
\affiliation{SRF Institute, Thomas Jefferson National Accelerator Facility, Newport News, Virginia 23606, USA.}
\author{Shaoheng Wang}
\affiliation{SRF Institute, Thomas Jefferson National Accelerator Facility, Newport News, Virginia 23606, USA.}

\date{\today}

\begin{abstract}
Superconducting radio-frequency (SRF) resonators with high intrinsic quality factors ($Q_0$) provide long photon lifetimes, enabling high-coherence quantum memories and strong coupling to quantum systems. Magnetically levitated oscillators reach similarly high mechanical quality factors because they eliminate mechanical clamping and minimize thermal-contact losses, making them attractive platforms for precision sensing and hybrid quantum architectures. In this work, we present an electromechanical architecture that couples a high-$Q$ superconducting cavity to a magnetically levitated oscillator. We develop a $5.4$~GHz SRF quarter-wave resonator, fabricated from bulk niobium and processed with state-of-the-art high-$Q$ surface treatments, that reaches an intrinsic quality factor of $Q_0 = 3.3\times10^{9}$ at $1.5$~K. A nominally identical cavity coated with Nb$_3$Sn sustains $Q_0 = 5\times10^{8}$ up to $4$~K, comparable to that of a bulk-niobium cavity operated at $2$~K, providing a pathway toward higher-temperature operation. We then model the integration of a $1$~mm diameter spherical magnet within the cavity and find that both the cavity resonance frequency and its loaded quality factor are highly sensitive to the oscillator displacement, and we identify stable levitation equilibria within the high-field region. Integrating an ultrahigh-coherence superconducting resonator with a levitated macroscopic oscillator opens a regime of cavity levitodynamics, providing new opportunities for magnetomechanics, quantum sensing, and precision measurement.
\end{abstract}

\maketitle

The quality factor ($Q$) is a fundamental figure of merit that quantifies $2\pi$ times the ratio of the energy stored in a resonator to the energy dissipated per RF cycle. In electromagnetic resonators, a high $Q$ corresponds to low energy dissipation and long photon lifetimes, allowing photons to remain stored for extended periods. In mechanical resonators, a high $Q$ reflects low mechanical damping, resulting in long-lived mechanical oscillations~\cite{pozar}.

Coupling high-coherence electromagnetic and mechanical modes remains a central frontier, particularly in quantum physics, circuit quantum electrodynamics, optomechanics, and electromechanics~\cite{aspelmeyer2014}. The strength of this coupling governs the ability to manipulate quantum states, providing a direct route toward ground-state cooling, coherent quantum-state transfer, and quantum-limited sensing~\cite{kokum}. Realizing these capabilities requires minimizing decoherence, motivating the continued development of high-$Q$ architectures~\cite{regal2011, dixit2021}. Extending these concepts to levitated mechanical objects has given rise to the field of levitodynamics~\cite{gonzalez2021}, in which eliminating mechanical supports offers new opportunities to achieve exceptionally low mechanical dissipation.

Magnetically levitated oscillators provide an attractive alternative to conventional clamped mechanical resonators by eliminating mechanical support and minimizing thermal-contact losses, thereby suppressing a dominant dissipation channel in macroscopic mechanical systems~\cite{gonzalez2021}. In superconducting magnetic levitation, a permanent magnet positioned above a type-II superconductor is levitated by diamagnetic screening currents, while pinned magnetic vortices provide the restoring forces required for stable levitation~\cite{hellman1988l}. Unlike optical trapping, superconducting magnetic levitation relies exclusively on static magnetic fields and is therefore free from photon-recoil heating and optical absorption~\cite{ahn2020}. Early demonstrations employed permanent magnets levitated above type-I and type-II superconductors with optical readout, achieving mechanical quality factors approaching $10^{4}$~\cite{timberlake2019,wang2019}. More recently, mechanical quality factors exceeding $10^{7}$ have been achieved, together with demonstrations of coupling to individual nitrogen-vacancy centers and motion detection using superconducting quantum interference device (SQUID)-based detectors~\cite{gieseler2020,vinante2020}, establishing magnetically levitated oscillators as a promising platform for hybrid quantum systems~\cite{gonzalez2021}. Furthermore, the combination of ultralow mechanical dissipation and cryogenic operation substantially suppresses thermomechanical noise, making magnetically levitated oscillators attractive for ultrasensitive force sensing, accelerometry, gravimetry, and magnetometry~\cite{amaral2025first, prat2017,goodkind1999,jackson2016,vinante2021, carney2025superconducting}.

\begin{figure*}[t]
\includegraphics[width=\textwidth]{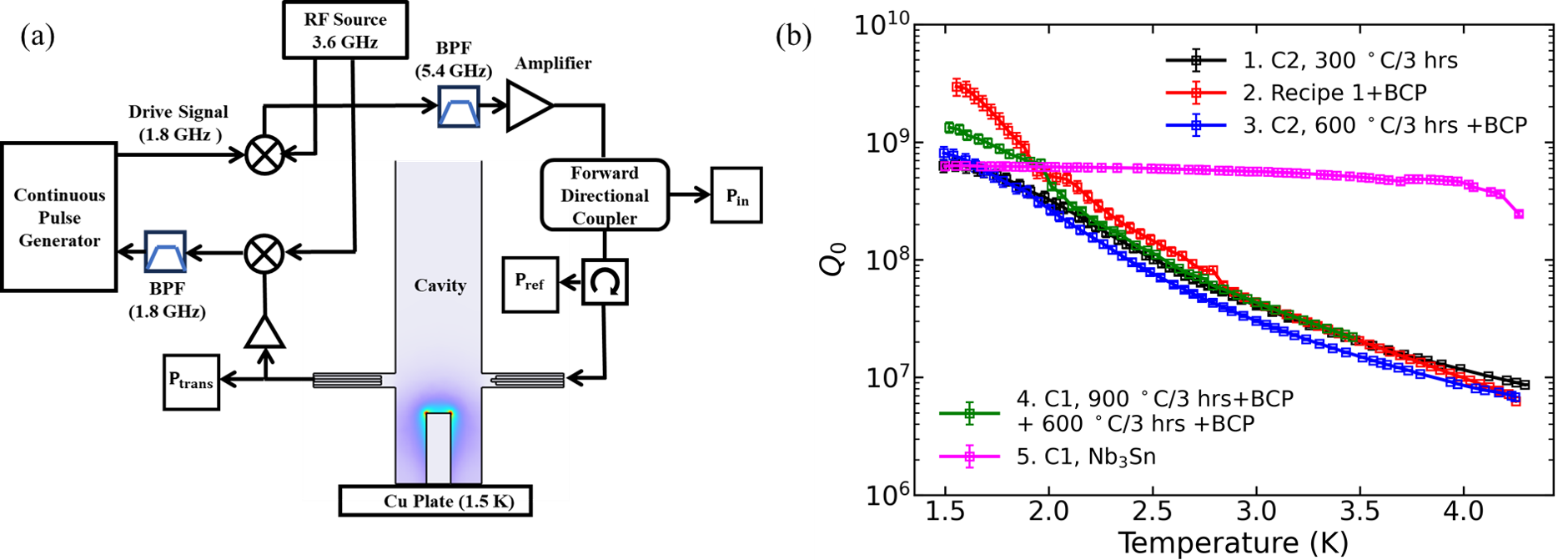}
\caption{\label{fig1}Experimental setup and cavity characterization. (a)~Schematic of the pulsed self-excited-loop microwave measurement system. (b)~Intrinsic quality factor $Q_0$ measured during cooldown from $4.5$ to $1.5$~K for Cavities 1 and 2 after each surface treatment. In all cases $Q_0$ increases as the BCS contribution to the surface resistance is suppressed with decreasing temperature.}
\end{figure*}

Three-dimensional elliptical superconducting radio-frequency (SRF) cavities have achieved ultrahigh quality factors ($Q \approx 10^{10}$--$10^{11}$) through advances in surface engineering, novel superconducting materials, and optimized cooling protocols~\cite{dhakal2013, grassellino2013, posen2020}. Beyond accelerator physics, superconducting microwave cavities are implemented as long-lived quantum memories capable of controlling, storing, and reading out quantum states~\cite{krasnok2024,belomestnykh2022}. However, the physical dimensions of conventional elliptical cavities limit their integration into compact tabletop platforms. These developments have recently been extended to compact quarter-wave resonators~\cite{takenaka2026,oriani2025}. Quarter-wave cavities combine small electromagnetic mode volumes with low dissipation by confining the fields within a compact geometry while minimizing losses at joints and dielectric interfaces. This compact footprint enables strong coupling to cavity fields without substantially degrading $Q$, making quarter-wave cavities well suited for quantum information processing~\cite{reagor2016, potts2020,tabuchi2015coherent}. The integration of high-coherence SRF resonators with the magnetically levitated oscillators described above, a regime we term cavity levitodynamics, remains largely unexplored.

Building on these advances, we investigate the coupled electromechanical dynamics that would arise from integrating our high-$Q$ superconducting quarter-wave resonator with an NdFeB magnet levitated 1~mm above the stub. Using experimental characterization, finite-element simulations, and analytical modeling, we map the full electromechanical response of the system and determine the conditions for stable levitation. This work establishes the foundation for cavity levitodynamics based on ultrahigh-coherence superconducting microwave resonators, offering a new platform for precision sensing and hybrid quantum devices.

Figure~\ref{fig1}(a) shows the experimental setup. The coaxial cavity comprises a cylindrical outer conductor of radius $7$~mm and height $36$~mm and an inner conductor (stub) of radius $2$~mm and height $12$~mm. The electric field is strongly localized around the stub, making the cavity mode highly sensitive to nearby mechanical perturbations~\cite{Kudra2020}.

We fabricated two nominally identical quarter-wave resonators from bulk niobium and applied a different sequence of surface treatments to each, measuring $Q_0$ after every step. After machining, both cavities received a $100~\mu$m buffered chemical polish (BCP) to remove the mechanically damaged layer~\cite{antoine2012}. Cavity 2 (C2) was baked at $300\,^\circ$C for 3~h (1), given an additional $10~\mu$m BCP (2), and then annealed at $600\,^\circ$C for 3~h followed by a further $10~\mu$m BCP (3). Cavity 1 (C1) was annealed at $900\,^\circ$C for 3~h and given a $10~\mu$m BCP, then annealed at $600\,^\circ$C for 3~h and given a final $10~\mu$m BCP (4). C1 was subsequently degreased, ultrasonically cleaned for 30~min, and coated on its inner surface with Nb$_3$Sn by vapor diffusion (5). The numbers in parentheses label the corresponding curves in Fig.~\ref{fig1}(b).

The cavity is mounted inside a vacuum chamber, thermally anchored to a copper cold plate, and cooled to $1.5$~K at a base pressure below $10^{-7}$~mbar. Two sensors attached to the cavity surface monitor its temperature throughout the measurement. To minimize magnetic-flux trapping during cooldown, three fluxgate magnetometers mounted outside the vacuum chamber record the ambient magnetic field, while a Helmholtz coil inside the Dewar actively suppresses the residual field to below $5$~mG. Two pin couplers terminate the coaxial lines $5$~mm above the stub, with the primary coupler overcoupled at $1.5$~K while the secondary coupler remains undercoupled.

To evaluate cavity performance, we employed a pulse-based self-excited loop (SEL), shown schematically in Fig.~\ref{fig1}(a). The cavity is driven by RF pulses generated within the loop at input powers ranging from milliwatts down to nanowatts. The transmitted signal is up-converted, amplified, and re-injected into the cavity. A directional coupler enables simultaneous monitoring of the forward and reflected powers, while a circulator isolates the amplification chain from the reflected signal. After the drive pulse is switched off on resonance, the loaded decay time ($\tau_L$) is determined by fitting the exponentially decaying reflected power. The loaded ($Q_L$) and external ($Q_{\text{ext}}$) quality factors are extracted from the incident ($P_{\text{in}}$), reflected ($P_{\text{ref}}$), and transmitted ($P_{\text{trans}}$) power profiles, yielding the intrinsic quality factor $Q_{0} = \omega \tau_{L}(1 + \beta_{1} + \beta_{2})$, where $\omega$ is the angular resonance frequency and $\beta_i = Q_0/Q_{\text{ext},i}$ are the coupling coefficients of the two couplers.

Figure~\ref{fig1}(b) shows the intrinsic quality factor ($Q_0$) measured during cooldown from $4.5$ to $1.5$~K for both resonators. In both cavities, $Q_0$ rises as the BCS contribution to the surface resistance is suppressed with decreasing temperature, and saturates below $\sim\!2$~K, where the temperature-independent residual resistance dominates. For Cavity 2, the initial $300\,^\circ$C, 3~h bake alone did not raise $Q_0$ above this residual limit (1). Adding a $10~\mu$m BCP after that bake improved $Q_0$ substantially (2), giving the best performance of this study, $Q_0 = 3.3\times10^{9}$ at $1.5$~K; the subsequent $600\,^\circ$C, 3~h anneal with a further $10~\mu$m BCP did not improve on this value (3). For Cavity 1, the $900\,^\circ$C, 3~h anneal with a $10~\mu$m BCP, followed by a $600\,^\circ$C, 3~h anneal with a final $10~\mu$m BCP, raised $Q_0$ at $1.5$~K to $1.4\times10^{9}$ (4). After Nb$_3$Sn coating, Cavity 1 sustained $Q_0 = 5\times10^{8}$ up to $4$~K (5), comparable to that of a bulk-niobium cavity operated at $2$~K. For all runs, $Q_{\text{ext},1}$ and $Q_{\text{ext},2}$ remained constant during cooldown, at $1.3\times10^{8}$ and $7.3\times10^{10}$, respectively. Over the full mW-to-nW drive range ($\gtrsim10^{12}$ stored photons), $Q_0$ showed no dependence on input power; at these photon numbers any two-level-system (TLS) loss is saturated.

\begin{figure}[t]
\includegraphics[width=\columnwidth]{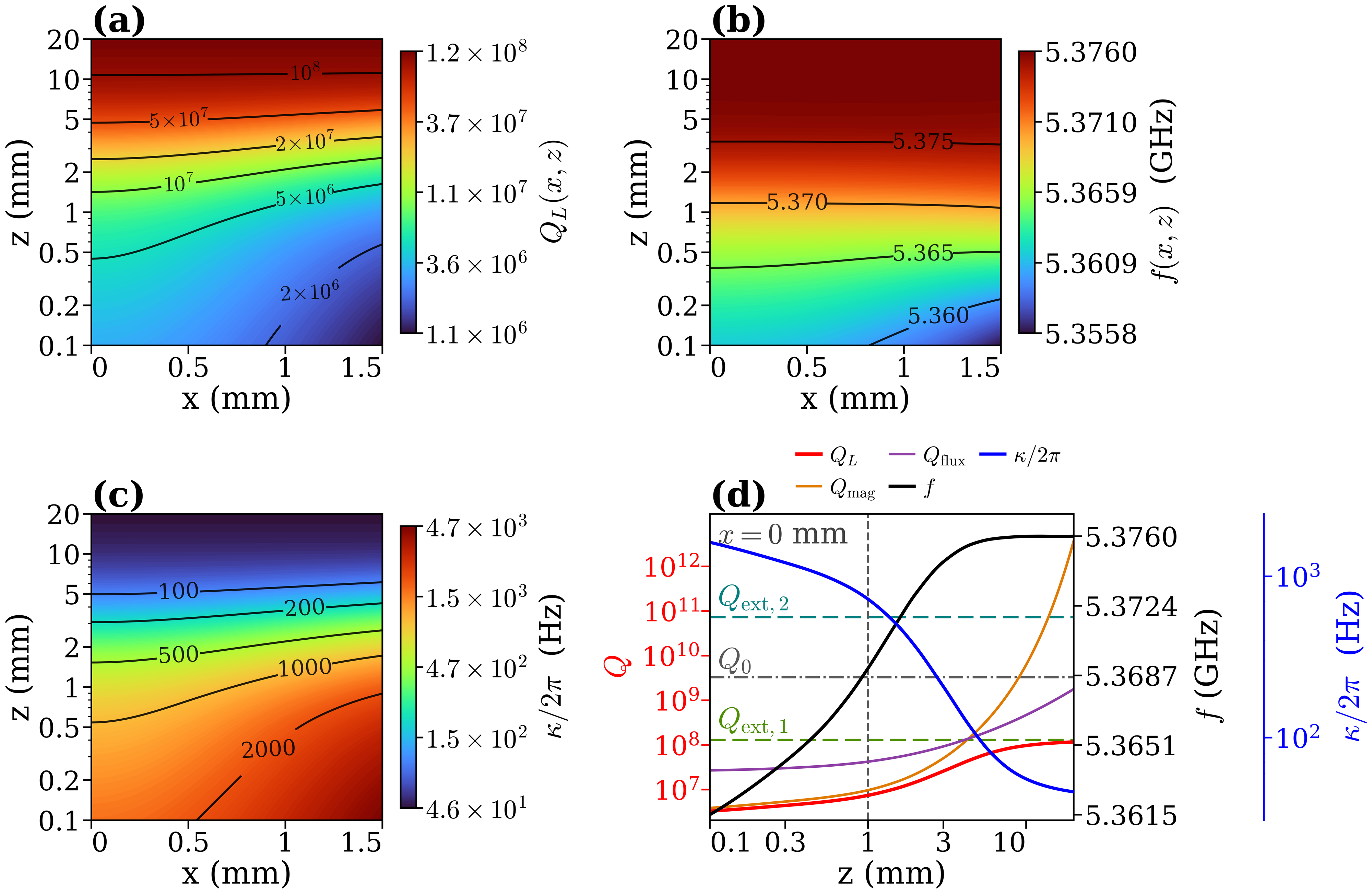}
\caption{\label{fig2}Electromechanical response of the cavity to the displacement of a $1$~mm diameter spherical NdFeB magnet with remanent flux density $B_r = 1.48$~T, field-cooled at lateral offset $x$ and surface-to-surface gap $z$ above the stub; all $z$ axes are logarithmic. (a) Loaded quality factor $Q_L$ of the cavity with the magnet and couplers, displaying three stages of degradation and recovery, set by RF currents in the magnet ($Q_{\text{mag}}$), flux trapped in the niobium ($Q_{\text{flux}}$), and the input coupler ($Q_{\text{ext},1}$). (b) Resonance-frequency shift, largest within the capacitive near field of the central post ($z < 3$~mm) and saturating at larger separations, tracking the rapid spatial decay of the cavity's electric field. (c) Total cavity damping rate $\kappa/2\pi = f_c/Q_L$, which reaches $4.7$~kHz near the stub and falls to $46$~Hz at $z=20$~mm. (d) Loaded quality factor $Q_L$ along the cavity axis with its loss channels, RF currents in the magnet ($Q_{\text{mag}}$) and trapped flux ($Q_{\text{flux}}$), and the $z$-independent $Q_0$, $Q_{\text{ext},1}$ and $Q_{\text{ext},2}$ (broken lines), together with the resonance frequency and damping rate.}
\end{figure}

Having established the baseline cavity performance, we next model the response of the high-$Q$ niobium cavity at $1.5$~K to a nearby permanent magnet. The sensitivity of niobium cavities to trapped magnetic flux has been studied extensively as a function of temperature, RF frequency, and applied magnetic field~\cite{bafia2025, dhavale2012}. We consider a commercially available, spherical NdFeB magnet with a diameter of $1$~mm and a remanent flux density of $B_r = 1.48$~T, corresponding to a peak field of $\tfrac{2}{3}B_r \simeq 0.99$~T at its poles. The magnet is displaced along the cavity axis above the stub. The cavity is cooled through its critical temperature in the presence of the magnet's field; throughout this paper, we refer to this process as field cooling.

Three mechanisms determine the cavity resonance frequency and quality factor. The electric field is concentrated at the stub, and hence the conducting magnet increases the local capacitance, acting as the movable plate of a parallel-plate capacitor. Consequently, the electromechanical frequency coupling is strongest near the stub and decreases rapidly as the magnet moves away~\cite{raut2021}. By contrast, magnet-induced losses in the niobium arise from the interaction between the magnet's static field and the superconducting surface~\cite{raut2022}. As the magnet approaches the surface, vortex entry and pinning increase the microwave losses by raising the surface resistance~\cite{Posen2023}. In addition, NdFeB is a normal conductor ($\rho \approx 1.5~\mu\Omega\,$m at room temperature), so RF currents induced in the magnet by the stub field dissipate energy within the magnet itself. An insulating magnet, such as ferrite, could avoid this loss while still allowing magnetic levitation, despite its lower magnetization, provided that its dielectric loss tangent at $1.5$~K is below approximately $10^{-4}$.

To quantify these effects, we model the coupled cavity--magnet system using a two-stage finite-element method (FEM). First, we calculate the static magnetic-field distribution generated by the permanent magnet. We then incorporate this field profile into the electromagnetic eigenmode simulation using a field-dependent impedance boundary condition that accounts for the spatial variation of the niobium's microwave surface resistance. The magnet is treated as a normal conductor with surface resistance $R_{s,\text{mag}} = \sqrt{\omega\mu_0\rho/2} \approx 0.18~\Omega$. This approximation is appropriate because the magnet's skin depth ($\approx 8~\mu$m) is much smaller than its radius. We use the room-temperature resistivity; since $R_{s,\text{mag}}\propto\sqrt{\rho}$ and the resistivity of the metallic magnet is not expected to increase on cooling, this choice gives an upper bound on the magnet loss. The power dissipated on each surface is calculated from the tangential RF magnetic field, $P = \tfrac{1}{2}\oint R_s |\mathbf{H}_t|^2\,dS$, and each loss channel is expressed as a quality factor, $Q_i = \omega_c U/P_i$, where $U$ is the stored energy. The baseline BCS ($R_{\text{BCS}}$) and residual ($R_{\text{i}}$) surface resistances are extracted from the measured intrinsic quality factor, $Q_0 = 3.3 \times 10^9$, using $R_s = G/Q_0$. Here, the geometry factor is $G = \SI{66}{\ohm}$, obtained from an eigenmode simulation of the cavity that includes all niobium surfaces. Here, $Q_0$ is the intrinsic quality factor of the unperturbed cavity, measured without the magnet. The magnet introduces two additional loss channels: flux trapped in the niobium ($Q_{\text{flux}}$) and RF currents induced in the magnet ($Q_{\text{mag}}$). The couplers contribute through their measured external quality factors. Thus, the loaded (total) quality factor is
\[
\frac{1}{Q_L} = \frac{1}{Q_0} + \frac{1}{Q_{\text{flux}}} + \frac{1}{Q_{\text{mag}}} + \frac{1}{Q_{\text{ext},1}} + \frac{1}{Q_{\text{ext},2}}.
\]

The local surface resistance of the niobium is determined by the magnetic field ($B$) at the superconducting surface. Because the magnet is field-cooled, magnetic flux is trapped wherever its field penetrates the niobium. This adds a field-dependent contribution to the surface resistance, $R_{\text{s}} = R_{\text{BCS}} + R_{\text{i}} + \alpha B$, with the $\alpha B$ term determining $Q_{\text{flux}}$. Here, $\alpha \approx 0.7~\text{n}\Omega/\text{mG}$ at $5~\text{GHz}$ and $1.5~\text{K}$~\cite{bafia2025}. The field exceeds the vortex-penetration threshold of $180~\text{mT}$ only for $z < 0.4$~mm. At fields above $400~\text{mT}$, the surface transitions to the normal-conducting state, resulting in a rapid increase in dissipation.

These regimes are reflected in the simulated quality factor as the magnet moves along the cavity axis [Figs.~\ref{fig2}(a) and \ref{fig2}(d)]. At levitation heights below approximately $4$~mm, dissipation is dominated by RF currents in the magnet, which exceed the vortex-induced losses in the niobium ($Q_{\text{mag}} < Q_{\text{flux}}$). Between approximately $4$ and $10$~mm, $Q_L$ recovers as the trapped field at the niobium surface decreases. Above approximately $10$~mm, $Q_L$ approaches its unperturbed value of $1.25\times10^{8}$, which is set by the input coupler. In contrast, the resonance-frequency shift is governed primarily by the cavity near field [Fig.~\ref{fig2}(b)]. The frequency sensitivity is greatest within approximately $3$~mm of the central stub, where the electric field is most strongly confined, and decreases substantially at larger separations. Figure~\ref{fig2}(c) shows the total cavity linewidth, or damping rate, $\kappa/2\pi = f_c/Q_L$. This linewidth decreases from approximately $4.7$~kHz near the stub to $46$~Hz at the largest separation computed ($z=20$~mm), slightly above the $43$~Hz of the unperturbed cavity because flux trapped in the niobium still carries about $7\%$ of the loss at this height.

At a levitation height of $1$~mm, the magnet is the dominant source of loss:
$Q_{\text{mag}} = 9.7\times10^{6}$ and $Q_{\text{flux}} = 4.2\times10^{7}$, compared with $Q_0 = 3.3\times10^{9}$. Thus, RF currents in the magnet carry $76\%$ of the total loss, and trapped flux $18\%$. Adding both couplers lowers the quality factor by a further $6\%$, to $Q_L = 7.4\times10^{6}$. The magnet therefore broadens the loaded linewidth from $43$~Hz to $\kappa/2\pi = f_c/Q_L = 723$~Hz [Fig.~\ref{fig2}(d)]. The cavity nevertheless remains highly responsive to the magnet: a $1~\mu$m displacement shifts the resonance by $5.46$~kHz, $7.6$ times the linewidth, with a fractional change in $Q_L$ of only $7\times10^{-4}$.

Figure~\ref{fig2}(d) also illustrates how the levitation height and input coupling determine $Q_L$. For $z\lesssim6$~mm, the quality factor without the couplers is lower than $Q_{\text{ext},1}$, indicating that magnet-induced losses, rather than external coupling, limit the loaded quality factor. At a height of $1$~mm, removing the primary input coupler would therefore increase $Q_L$ only modestly, because the magnet-induced internal losses would remain dominant. Increasing the levitation height reduces these losses and raises $Q_L$, while weakening the input coupling (by using a shorter coupler pin) further increases $Q_L$ at larger heights. These improvements involve two trade-offs: the frequency shift per unit displacement decreases as the levitation height increases, and weaker input coupling reduces the fraction of the cavity energy decay that occurs through the input port, $\kappa_{\text{ext},1}/\kappa = Q_L/Q_{\text{ext},1}$. Optimizing the operating point therefore requires balancing the cavity linewidth, frequency sensitivity, and signal extraction.

\begin{figure}[t]
\includegraphics[width=\columnwidth]{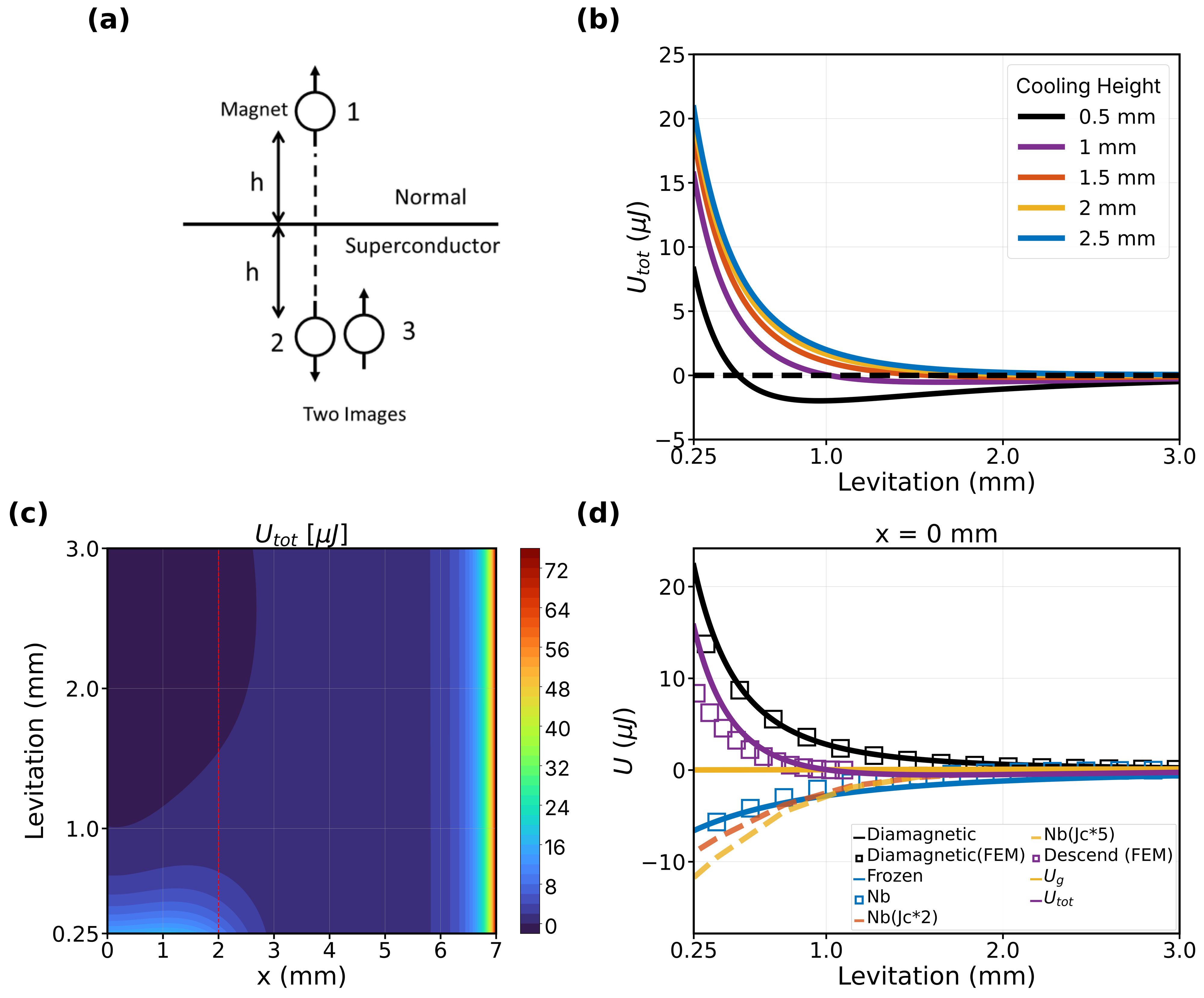}
\caption{\label{fig3} Potential energy analysis of the cavity--magnet system. (a) Schematic of the frozen-image model. A magnet field-cooled at a height $h$ above a type-II superconductor (1) is represented by a static frozen-flux image (3), originating from vortex pinning, and a dynamic diamagnetic image (2), corresponding to Meissner flux expulsion. (b) Total potential energy versus field-cooling height $h$. (c) Calculated potential energy landscape for $h = 1$~mm, with heights measured from the plane of the stub top for all $x$; the energy varies continuously across the stub edge (dashed line) and rises towards the cavity wall. (d) Vertical potential energy profile for a magnet cooled at $h = 1$~mm. The hysteretic response during approach and withdrawal illustrates the role of flux pinning, which establishes a stable levitation equilibrium.}
\end{figure}

The potential energy of the cavity--magnet interaction is calculated using a frozen-image model~\cite{Hull1999, Cansiz2003}, extended here to the cavity geometry with images in the stub, the floor and the cavity wall. As illustrated in Fig.~\ref{fig3}(a), a magnet field-cooled at a height $h$ above a type-II superconductor is represented by two effective image sources: a static frozen-flux image and a dynamically varying diamagnetic image. The frozen-flux image, pinned at a depth $-h$ corresponding to the field-cooling position (where the magnet is held while the niobium cools through $T_c$), represents the magnetic flux pinned within the superconductor. Its interaction with the magnet provides the restoring forces required for stable levitation. In contrast, the diamagnetic image follows the instantaneous position of the magnet and produces the repulsive Meissner contribution.

The total potential energy of the cavity--magnet system is obtained by summing the frozen-flux, $U_{\text{frozen}}(x,y,z,\theta,\phi)$, and diamagnetic, $U_{\text{dia}}(x,y,z,\theta,\phi)$, image-interaction energies together with the gravitational potential $Mgz$, with
\begin{equation}
\label{eq:udia}
U_{\text{dia}} = \frac{\mu_{0}m_{1}m_{2}}{4\pi} \left[ \frac{1+\cos^{2}\theta}{(2z)^{3}} \right]
\end{equation}
\begin{equation}
\label{eq:ufrozen}
U_{\text{frozen}} = \frac{\mu_{0}m_{1}m_{2}}{4\pi} \left[ \frac{A(x,y,z,h,\theta,\phi)}{(x^{2}+y^{2}+(z+h)^{2})^{5/2}} \right]
\end{equation}
\begin{multline}
\label{eq:A}
A(x,y,z,h,\theta,\phi) = x^{2}(1-3\sin^{2}\theta\cos^{2}\phi) \\
+ y^{2}(1-3\sin^{2}\theta\sin^{2}\phi) + (z+h)^{2}(1-3\cos^{2}\theta) \\
- 6xy\sin^{2}\theta\cos\phi\sin\phi - 6x(z+h)\sin\theta\cos\theta\cos\phi \\
- 6y(z+h)\sin\theta\cos\theta\sin\phi
\end{multline}
Here $m_{1}$ is the dipole moment of the magnet and $m_{2}$ that of the corresponding image, whose sign fixes the repulsive (diamagnetic) or attractive (frozen-flux) character of each term; the coordinates $(x,y,z)$ describe the translational position of the magnet, with $z$ measured from the plane of the stub top for all $x$, $h$ is the field-cooling height, and $(\theta,\phi)$ specify the rotational orientation of the dipole. As in Fig.~\ref{fig2}, $z$ and $h$ are surface-to-surface gaps between the magnet and the plane of the stub top; in Eqs.~\eqref{eq:udia}--\eqref{eq:A} the dipole sits at the magnet centre, so $z$ and $h$ enter as $z+a$ and $h+a$, where $a=0.5$~mm is the magnet radius. The diamagnetic image of the stub is switched off smoothly across the stub edge, over a width set by the height of the magnet.

We calculate the total potential energy as a function of vertical displacement for several field-cooling heights, with the magnet centered on the cavity axis. As shown in Fig.~\ref{fig3}(b), the interplay between Meissner repulsion and vortex pinning produces a stable trapping potential whose depth and stiffness are highly sensitive to the field-cooling height $h$. For small $h$, strong flux pinning yields pronounced axial and transverse restoring forces, whereas increasing the cooling height reduces the pinning strength and weakens the confinement. Figure~\ref{fig3}(c) shows the potential energy landscape for a cooling height of $1$~mm, in which the rise in energy beyond the stub edge and towards the cavity wall encloses a local minimum on the cavity axis.

For a magnet field-cooled at a $1$~mm gap above the stub, the oscillation frequencies follow from the eigenvalues of the stiffness matrix $K = \nabla^{2}U$ evaluated at the equilibrium position, which lies $10~\mu$m below the field-cooling height. Assuming cylindrical symmetry, which eliminates the $y$ and $\phi$ components, the ratio of vertical to lateral frequency reduces to~\cite{Cansiz2003}
\begin{equation}
\label{eq:fratio}
\frac{f_{z}}{f_{x}} = \left[ \frac{1+\cos^{2}\theta}{16\cos\theta}\left(1+\frac{h}{z}\right)^{5} - 2 \right]^{1/2}
\end{equation}
which equals $\sqrt{2}$ for an axial dipole at $z = h$ and $1.45$ at the equilibrium height. Diagonalizing the full stiffness matrix for our magnet, of mass $M = 3.9$~mg and moment $m = 6.2 \times 10^{-4}~\text{A\,m}^{2}$, gives a vertical mode $f_{z} = 157$~Hz, a lateral mode $f_{x} = 53$~Hz, and a librational mode $f_{\theta} = 319$~Hz. The lateral and librational motions are mixed by the cross-stiffness $k_{x\theta}$, which softens $f_{x}$ from its uncoupled value of $106$~Hz and stiffens $f_{\theta}$ from $255$~Hz, so that Eq.~\eqref{eq:fratio} applies to the uncoupled stiffnesses.

We compare the analytical result against a finite-element calculation in which the superconductor's reaction field is obtained from a phenomenological critical-state description of the current density $J_c(B)$, implemented in the $H$--$\phi$ formulation of Refs.~\cite{arsenault2020, grilli2018}. That formulation has been benchmarked against measured levitation force--displacement curves, including their field-cooling hysteresis, for magnets above bulk high-$T_c$ superconductors~\cite{sass2015h}, reproducing both the equilibrium position and the loop area without free parameters. We apply the same framework to niobium, using $J_c(B)$ for fine-grain niobium from measurements in Ref.~\cite{dhavale2012} at $1.5$~K. As shown in Fig.~\ref{fig3}(d), the FEM results agree well with the frozen-image model for separations $z > 0.25$~mm, with deviations at smaller distances arising from the breakdown of the point-dipole approximation. Across the parameter space, the levitation response falls into three regimes. In the Meissner-dominated regime, the interaction is primarily repulsive, with negligible flux pinning at large separations. In the flux-pinning regime, trapped vortices produce hysteretic restoring forces that depend on the field-cooling history. In the stable levitation regime, near the field-cooling height, the balance of Meissner repulsion and vortex pinning establishes a well-defined equilibrium with restoring forces for both positive and negative displacements. Increasing the critical current density $J_c$ strengthens the flux-pinning force, enabling the stable levitation of larger or more strongly magnetized objects at greater levitation heights. These results suggest that superconductors with higher critical current densities, such as Nb$_3$Sn, could further improve the levitation performance of the proposed cavity architecture.

Figures~\ref{fig2} and \ref{fig3} describe, respectively, how the magnet modifies the cavity's RF response and how it moves in its trap. We now combine the two to examine the coupling between the cavity mode and the mechanical motion: we quantify the interaction strength, evaluate how sensitively the cavity detects the motion of the levitated magnet, and assess the requirements for preparing the coupled system in the quantum regime.

The electromechanical interaction is characterized by the vacuum optomechanical coupling rate
\begin{equation}
g_{0} = \left(\frac{\partial \omega_{c}}{\partial z}\right)x_{\mathrm{zpf}},
\end{equation}
where \(x_{\mathrm{zpf}}=\sqrt{\hbar/(2M\Omega_{z})}\) is the zero-point displacement of the vertical mode~\cite{aspelmeyer2014}. The rate \(g_{0}\) is the cavity-frequency shift associated with a mechanical displacement equal to the zero-point amplitude \(x_{\mathrm{zpf}}\), which sets the single-photon optomechanical coupling strength. For the \(M=3.9\)~mg magnet field-cooled at a \(1\)~mm gap above the stub, the frequency pull \(\partial f_{c}/\partial z=5.46\)~kHz/\(\mu\)m and the vertical-mode frequency \(\Omega_{z}/2\pi=157\)~Hz yield \(x_{\mathrm{zpf}}=1.2\times10^{-16}\)~m, near the macroscopic-mirror end of optomechanical systems~\cite{kippenberg2008}, and \(g_{0}/2\pi=0.64~\mu\)Hz. Although \(g_{0}\) is intrinsically small owing to the macroscopic mass of the oscillator, the low cavity loss partially compensates for this limitation. At this height, the loaded linewidth is \(\kappa/2\pi=723\)~Hz, dominated by magnet-induced losses, compared with \(43\)~Hz for the unperturbed cavity (\(1.6\)~Hz intrinsic). The resulting ratio \(g_{0}/\kappa=8.8\times10^{-10}\) is more than two orders of magnitude larger than that inferred for the vertical mode in flux-mediated readout of a magnetically levitated superconductor~\cite{schmidt2024}. The displacement of the magnet also changes the cavity linewidth, giving the interaction a dissipative component, \(\partial\kappa/\partial z\), in addition to the dispersive coupling \(\partial\omega_{c}/\partial z\)~\cite{elste2009,aspelmeyer2014}. At the \(1\)~mm gap, the fractional change in \(Q_L\) of \(7\times10^{-4}\) per micrometer gives \(|(\partial\kappa/\partial z)/(\partial\omega_{c}/\partial z)|\approx1\times10^{-4}\). The coupling is therefore dispersive, and the dissipative contribution is neglected.

In our setup, the magnet sits in the strong electric field above the stub, so its vertical motion shifts the cavity frequency by \(5.46\)~kHz per micrometer. This modulation can be read out through the same input coupler used to characterize the cavity. The vertical mode lies within the cavity linewidth, allowing a resonant probe to follow the motion directly. On resonance, the motion modulates, to linear order, only the phase of the reflected field, leaving the intracavity amplitude, and hence the force on the magnet, unchanged. Thus, the probe reads out the motion without dynamical back-action~\cite{kippenberg2008}. A resonant probe of input power \(P_{\mathrm{in}}\) sustains \(n_{d}=4\kappa_{\mathrm{ext}}P_{\mathrm{in}}/(\hbar\omega_{c}\kappa^{2})\) intracavity photons, and the single-sided displacement imprecision at the mechanical frequency is~\cite{clerk2010,teufel2009}
\begin{equation}
S_{zz}^{\mathrm{imp}}(\Omega_{z})=\frac{\kappa}{8\eta n_{d}G^{2}}\left(1+\frac{4\Omega_{z}^{2}}{\kappa^{2}}\right)\left(1+2n_{\mathrm{add}}+2\bar{n}_{c}\right),
\end{equation}
where \(G=\partial\omega_{c}/\partial z\), \(\eta=\kappa_{\mathrm{ext}}/\kappa\) is the detection efficiency, set by the fraction of the signal that leaves through the input coupler, \(n_{\mathrm{add}}\approx10\) is the noise added by a cryogenic amplifier, and \(\bar{n}_{c}=[\exp(\hbar\omega_{c}/k_{B}T)-1]^{-1}\) is the thermal occupation of the cavity. A probe power of \(0.5~\mu\)W, corresponding to \(n_{d}\simeq6.9\times10^{12}\), with \(\eta=6\%\) and \(\bar{n}_{c}\approx5.3\) at \(1.5\)~K, gives \([S_{zz}^{\mathrm{imp}}]^{1/2}\approx7\times10^{-15}\)~m/\(\sqrt{\mathrm{Hz}}\). The single-sided thermal motion of the magnet on resonance~\cite{aspelmeyer2014},
\begin{equation}
S_{zz}^{\mathrm{th}}(\Omega_{z})=\frac{4k_{B}TQ_{m}}{M\Omega_{z}^{3}},
\end{equation}
gives \([S_{zz}^{\mathrm{th}}]^{1/2}\approx1.5\times10^{-11}\)~m/\(\sqrt{\mathrm{Hz}}\) for the mechanical quality factor \(Q_{m}=10^{4}\) used below, more than three orders of magnitude above the imprecision. This resolution is also robust against changes in the cavity tuning: varying the levitation height over \(z=0.25\)--\(10\)~mm and \(Q_{\text{ext},1}\) between its present value and \(Q_0\) changes \(Q_L\) from \(4.2\times10^{6}\) to \(3.4\times10^{8}\) [Fig.~\ref{fig2}(d)], while the same \(0.5~\mu\)W probe maintains an imprecision at least \(20\) times below the thermal motion.

Preparing quantum states of the coupled system, however, requires more than resolving the motion. A drive on the red mechanical sideband, \(\omega_{d}=\omega_{c}-\Omega_{z}\), enhances the coupling to \(g=g_{0}\sqrt{n_{d}}\) and transfers phonons from the magnet's oscillation into the cavity through anti-Stokes scattering, thereby cooling the mechanical mode~\cite{wilson2007theory,kippenberg2008}. In contrast, for a drive on the blue sideband, Stokes scattering dominates, adding phonons and amplifying the motion~\cite{kippenberg2005}.

For a conservative mechanical quality factor \(Q_{m}=10^{4}\), measured for magnets levitated above superconductors~\cite{timberlake2019,wang2019}, and a damping rate \(\Gamma_{m}=\Omega_{z}/Q_{m}\), the classical cooperativity \(C=4g^{2}/(\kappa\Gamma_{m})\)~\cite{aspelmeyer2014,teufel2011} reaches unity at the same \(n_{d}\simeq6.9\times10^{12}\) used for the readout. Accounting for the cavity response at the drive frequency, this photon number requires an input power of
\begin{equation}
P_{\mathrm{in}}=n_{d}\hbar\omega_{d}\,\frac{(\kappa/2)^{2}+\Omega_{z}^{2}}{\kappa_{\mathrm{ext}}}\simeq0.6~\mu\mathrm{W}
\end{equation}
through our input coupler (\(\kappa_{\mathrm{ext}}/2\pi=41\)~Hz). Ground-state cooling, however, requires the quantum cooperativity \(C_{q}=C/\bar{n}_{\mathrm{th}}\) to exceed unity~\cite{aspelmeyer2014,teufel2011}, where \(\bar{n}_{\mathrm{th}}=k_{B}T/(\hbar\Omega_{z})\simeq2.0\times10^{8}\) at \(1.5\)~K. At a \(1\)~mm gap, the magnet-broadened linewidth places the system in the unresolved-sideband regime (\(\Omega_{z}/\kappa=0.22\), with a quantum back-action cooling limit of \(\bar{n}_{\min}=(\kappa/4\Omega_{z})^{2}\simeq1.3\)~\cite{aspelmeyer2014}). Achieving \(C_{q}=1\) would require approximately \(10^{2}\)~W, and even then the phonon occupation could not fall below this limit. The cavity field introduces a second limitation: phonons are transferred to the cavity mode, whose thermal occupation of \(\bar{n}_{c}\approx5.3\) photons at \(1.5\)~K bounds the achievable phonon occupation. Ground-state cooling therefore also requires a colder photon bath: the thermal occupation of the \(5.4\)~GHz mode falls below unity only below approximately \(0.37\)~K, and substantially lower temperatures, for example, in a dilution refrigerator, are needed to suppress it.

A further limitation arises from the drive itself: the cavity converts phase noise of the detuned microwave drive into amplitude fluctuations of the intracavity field, thereby producing a fluctuating force on the magnet~\cite{aspelmeyer2014}. Ground-state cooling requires the symmetrized power spectral density of the source frequency noise at the mechanical frequency to satisfy \(\bar{S}_{\omega\omega}(\Omega_{z})\lesssim g_{0}^{2}\hbar Q_{m}/(k_{B}T)\)~\cite{rabl2009,aspelmeyer2014}, so it also requires a larger \(g_{0}\).

Despite these limitations, the intrinsic linewidth of our cavity points to a much more favorable regime. Without magnet-induced losses, the cavity at critical coupling would reach \(\Omega_{z}/\kappa=48\), comparable to state-of-the-art electromechanical~\cite{teufel2011} and optomechanical~\cite{hofer2010} systems. Even in this regime, achieving \(C_{q}=1\) would require approximately \(2.1\)~W for \(Q_{m}=10^{4}\), a power that becomes independent of \(\kappa\) in the resolved-sideband limit at critical coupling, or approximately \(2\)~mW for \(Q_{m}=10^{7}\)~\cite{gieseler2020,vinante2020}. The intrinsic quality factor demonstrated here, \(Q_{0}=3.3\times10^{9}\), therefore provides a promising foundation for reaching the resolved-sideband regime~\cite{schliesser2008resolved} and, ultimately, the quantum regime, provided that magnet-induced dissipation is suppressed through higher-\(J_{c}\) surfaces such as Nb$_{3}$Sn, larger levitation heights, or smaller magnetic moments, together with further improvements in \(Q_{0}\) and \(Q_{m}\).

In summary, we have demonstrated a compact, 5.4~GHz superconducting radio-frequency niobium quarter-wave resonator with an intrinsic quality factor of $3.3\times10^{9}$ at $1.5$~K. This value is among the highest reported for a niobium coaxial resonator~\cite{takenaka2026,oriani2025} and was achieved well above millikelvin temperatures. A nominally identical cavity coated with Nb$_3$Sn sustains a quality factor at $4$~K comparable to that of the high-$Q$-treated niobium cavity at $2$~K. This result points toward ultrahigh-coherence operation at elevated temperatures while retaining the compact geometry required for tabletop quantum experiments.

By modeling the integration of a macroscopic, magnetically levitated NdFeB oscillator with this ultralow-loss cavity, we propose an electromechanical platform in which both the cavity resonance frequency and the loaded quality factor are predicted to be highly sensitive to the oscillator's displacement. Finite-element simulations and frozen-image modeling identify stable levitation equilibria within the cavity's capacitive near field, where electromechanical coupling remains strong despite magnet-induced losses. At a loaded quality factor of $7.4\times10^{6}$, a displacement of $1~\mu$m shifts the resonance by $7.6$ linewidths. This combination of high-$Q$ SRF cavities and magnetic levitation opens opportunities for magnetomechanics, quantum sensing, precision force measurements, and future quantum-transduction experiments.

\begin{acknowledgments}
We thank Francesco Grilli, Alexandre Arsenault, and Felipe Sass for fruitful discussions regarding a phenomenological model for the superconductor--magnet interaction. This material is based upon work supported by the U.S. Department of Energy, Office of Science, Office of Nuclear Physics under Contract No. 89243126CSC000213 and was also supported by the Laboratory Directed Research and Development (LDRD) program at Thomas Jefferson National Accelerator Facility. Michael E. Tobar was supported by the Australian Research Council Centre of Excellence for Dark Matter Particle Physics (CE200100008).
\end{acknowledgments}

\bibliography{apssamp}

\end{document}